\documentclass[multphys,vecphys]{svmult}
\usepackage[latin2]{inputenc}
\newcommand{\be}{\begin{equation}}
\newcommand{\ee}{\end{equation}}
\usepackage{amsfonts}
\usepackage{amssymb}
\usepackage{natbib}
\usepackage{makeidx}
\usepackage{graphicx}

\usepackage{multicol}
\usepackage[bottom]{footmisc}

\makeindex

\begin{document}

\title*{G7 country Gross Domestic Product (GDP) time correlations. \\
A graph network analysis}
\titlerunning{G7 correlation. A graph network analysis.}
\author{J. Mi\'skiewicz\inst{1} \and
M. Ausloos\inst{2}}
\institute{Institute of Theoretical Physics, University of Wroc\l{}aw, pl. M.
Borna 9, 50-204 Wroc\l{}aw, Poland
\texttt{jamis@ift.uni.wroc.pl}
\and SUPRATECS, B5, University
of Li$\grave e$ge, B-4000 Li$\grave e$ge, Euroland
\texttt{marcel.ausloos@ulg.ac.be}}

\maketitle

\section{Introduction}
The G7 countries (France, USA, United Kingdom, Germany, Japan, Italy, Canada) are the most developed countries  in the world, but such statement leaves unanswered the question on which of those is the most important one and of course what kind of dependencies exists between them. Of course this subject has been considered along various lines of analysis \cite{Frankel}, which usually require a detailed knowledge of the analysed objects and therefore are difficult to pursue. Our own question is to investigate the dependence and leadership problem on a very limited number of data. Within this paper correlations between G7 countries, are investigated on the basis of their Gross Domestic Product (GDP). GDP is one of the most important parameters describing state of an economy and is extensively studied (Lee et~al. 1998, Ormerod 2004).

The annual GDP records\footnote{http:$ \slash \slash$www.ggdc.net$ \slash $index-dseries.html\#top}, considered as a discrete time series are used over the last 53 years (since 1950 till 2003) in order to evaluate GDP increments and distances between those countries. Different distance functions are used and the results compared. Distance matrices are calculated in the case of discrete Hilbert spaces $ L_q $ $ (q=1,2) $, Eq. (\ref{discrete}), a statistical correlation distance, Eq. (\ref{stat}), and a difference between increment distributions, Eq. (\ref{distr_dist}). The distance functions were chosen here below taking into account considerations on basic properties of the data. The distance matrices are then analysed using graph methods in the form of a unidirectional or bidirectional chain  (UMLP and BMLP respectively) \cite{bidirectional} as well as through the locally minimal spanning distance tree (LMST).
\nopagebreak

\section{Distance and graph analysis}

In the case of discrete time series the metrics can be defined in the Hilbert space $ L_q $ $(q=1,2)$ in a standard way \cite{analiza}
\begin{equation}
\label{discrete}
d_q (A,B) = ( \sum_{i=1}^n |a_i - b_i |^q)^{\frac{1}{q}} ,
\end{equation}
 where $ A,B $ are time series: $ A=(a_1,a_2,\ldots , a_n) $, $ B=(b_1,b_2, \ldots , b_n) $.
The statistical correlation distance is used in the form:
\begin{equation}
\label{stat}
d(A,B)_{(t,T)} = \sqrt{\frac{1}{2}(1- corr_{(t,T)} (A,B))},
\end{equation}
where $ t $ and $ T $ are the final point and the size of the time window over which an average is taken respectively; the correlation function is defined as:
\begin{equation}
corr_{(t,T)} (A,B) = \frac{<A B>_{(t,T)} - <A>_{(t,T)} <B>_{(t,T)} }{\sqrt{(<A2>_{(t,T)} - <A>^2_{(t,T)}) (<B2>_{(t,T)} - <B>^2_{(t,T)}})} .
\end{equation}
The brackets $ < $ ... $> $ denote a mean value over the time window $ T $ at time $ t $.

Additionally the distribution $ p(r) $ function of GDP yearly increments $ (r) $ is evaluated and the correlations between countries are investigated using $ {\cal L}_q $ $ (q=1) $ metrics \cite{analiza}
\begin{equation}
 \label{distr_dist}
 d_{{\cal L}_q} (A,B) =  [\int_{-\infty}^{+\infty} |p_A (r) - p_B(r )|^q d r]^{\frac{1}{q}} .
\end{equation}
Since the statistical parameters describing GDP increments are very close to the normal distribution \cite{bidirectional} it is assumed that this distribution well describes the GDP increments distribution.

There are different advantages to each of those distance functions. The discrete Hilbert space  $ L_q $ distance Eq. (\ref{discrete}) can be applied to any data and does not require any special properties of the data so this method seems to be very useful for comparing various sets of data. The second method Eq. (\ref{stat}), a statistical distance, is specially sensitive to linear correlations. The third method Eq. (\ref{distr_dist}) is the most sophisticated one since it requires a knowledge of the data distribution function, but then points out to similarities between data statistical properties. The main disadvantage of the last method is that it is sensitive to the size of the data set, since it is based on the whole distribution function.

The distance matrices are built in a varying size time window moving along the time axis.
The distance matrices are analysed by network methods - in the form of LMST and correlation chains (CC).
The topological properties of such trees and graphs, generated as a visualisation of the correlation between GDP in G7 countries allow us to gain some practical information about the weakest points of the networks and some possible roots for crashes, recessions or booms as will be investigated in details in a following paper.

Our present analysis focuses on the globalization process of G7 country economies, which is understood as an increasing resemblance between development patterns. The question is investigated by means of the total graph weight which is defined as a sum of distances between the countries for a given graph type (for LMST) and the mean distance for CC.
LMST is a modification of the Minimum Spanning Tree algorithm \cite{mst}. It is built under the constraints that the initial pair of nodes on the tree are the countries with the strongest correlation between their GDP. CC are investigated in two forms: unidirectional and bidirectional minimum length chains (called UMLP and BMLP respectively) \cite{bidirectional}.
UMLP and BMLP algorithms are simplifications for LMST, where the closest neighbouring countries are attached at the end of a chain. In the case of the unidirectional chain the initial node is an arbitrary chosen country. Therefore in the case of UMLP the chain is expanded in one direction only, whereas in the bidirectional case countries might be attached at one of both ends depending on the distance value.

Moreover a percolation threshold is defined as the distance value at which all countries are connected to the network. The percolation threshold has been investigated for the different distance measures. This technique allows us to observe structures in GDP relationships between countries.

%\section{Graph analysis}

\section{Results}
\begin{figure}
    \includegraphics[scale=0.4,angle=-90]{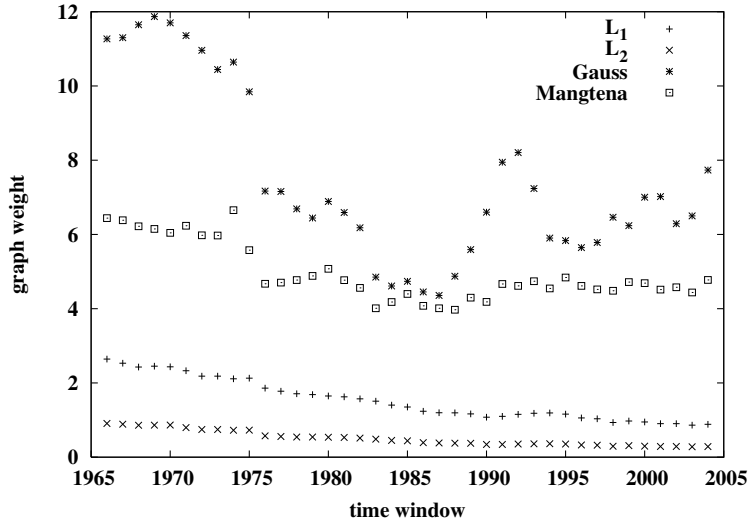}
    \caption{The time evolution of the graph weight for different distance measures. \newline The time window size is equal to 15 yrs.}
    \label{fig:tree}

\end{figure}

\begin{figure}
    \includegraphics[angle=-90,scale=0.25]{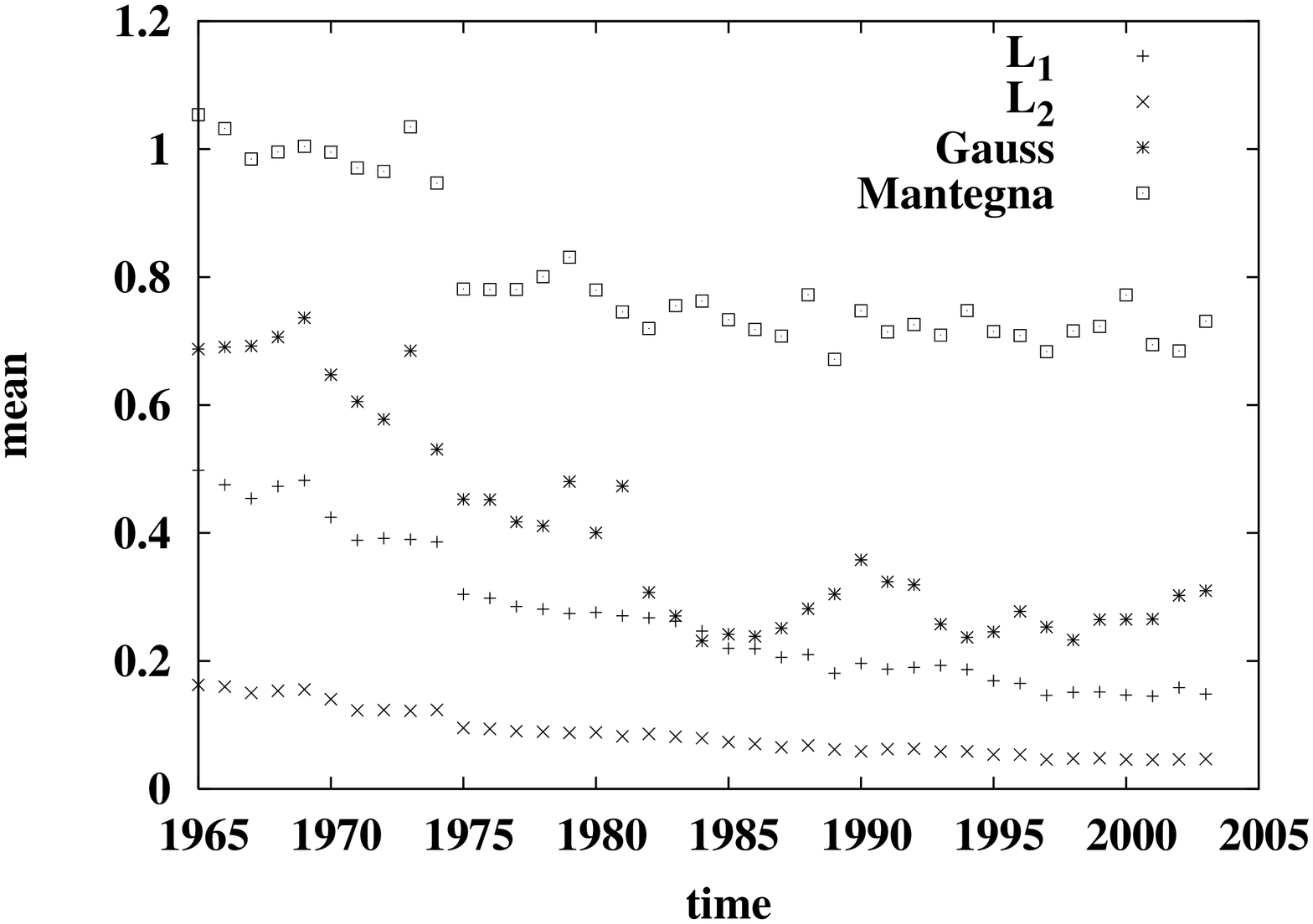}
    \includegraphics[angle=-90,scale=0.25]{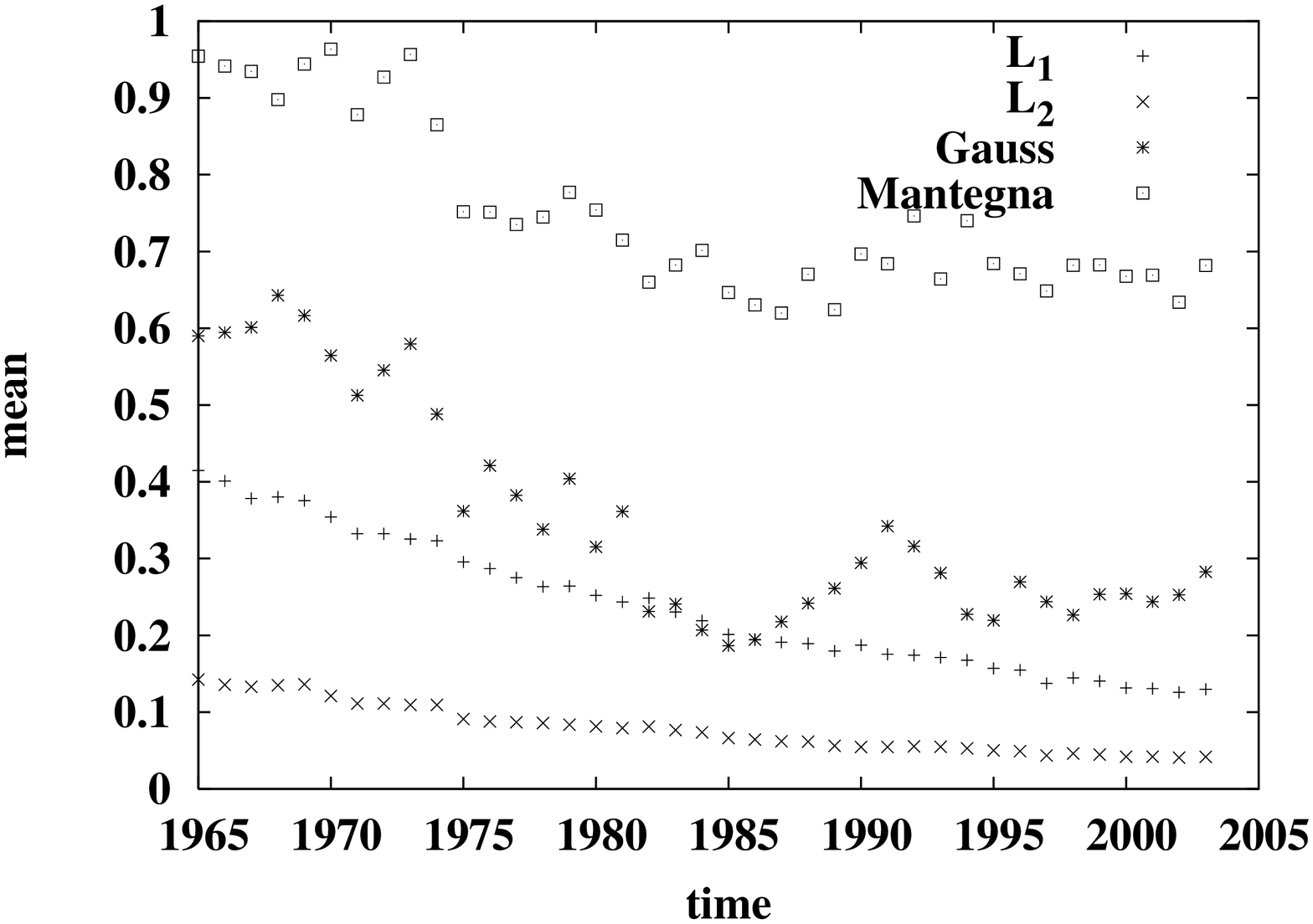}
    \caption{The time evolution of the total length of uni- and bidirectional chains for different measures. The time window size is equal to 15 yrs. }
    \label{fig:all}
\end{figure}

\begin{figure}
    \includegraphics[angle=-90,scale=0.25]{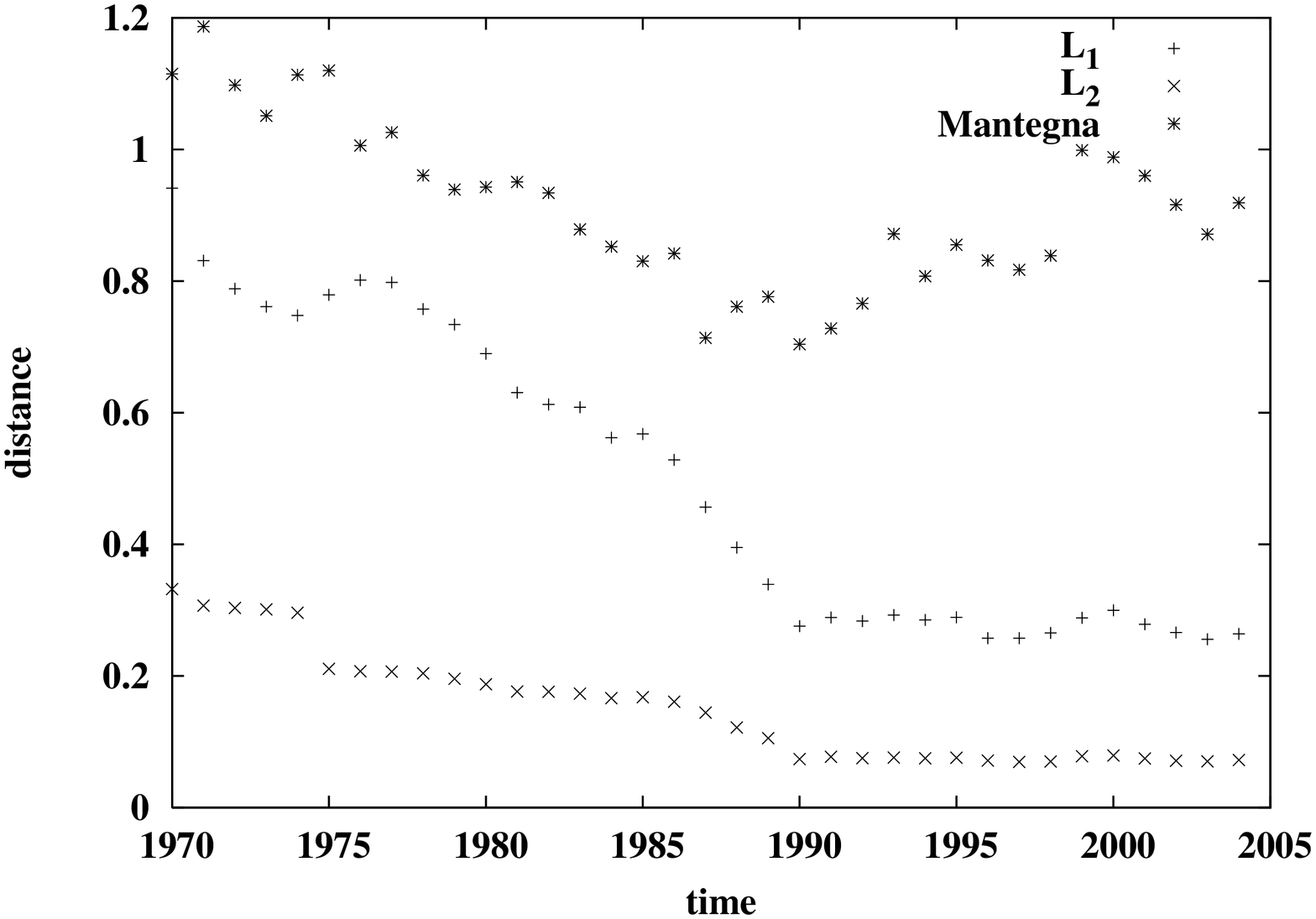}
    \includegraphics[angle=-90,scale=0.25]{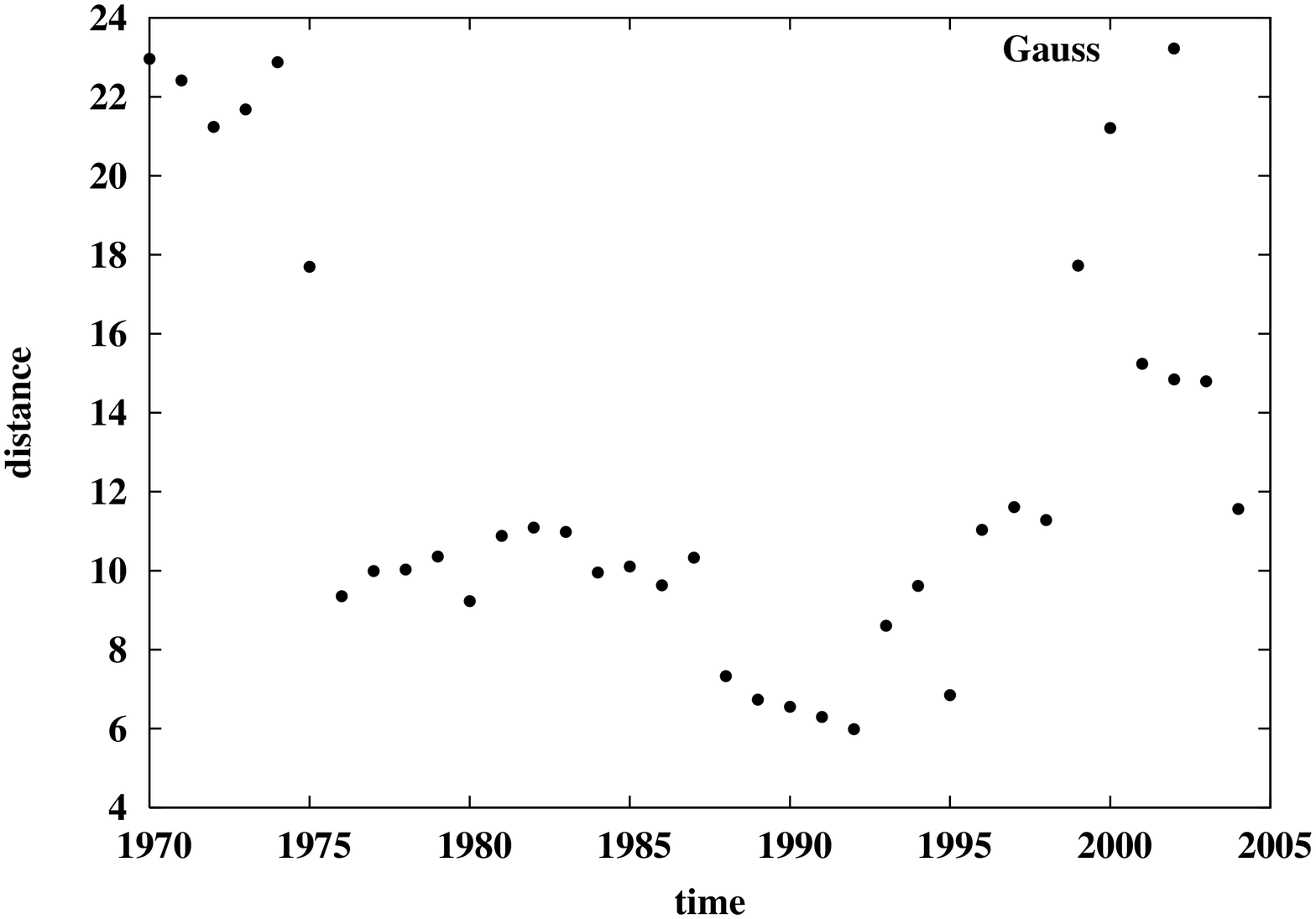}
    \caption{The time evolution of percolation threshold for different measures. The time window size is equal to 15 yrs.}
    \label{fig:perkol}
\end{figure}

The analysis is discussed here for a 15 years time window, which allows to observe the globalization process and statistically compare results obtained by different methods. The graph and percolation analysis were performed in the case of $L_1$, $L_2$, ${\cal L}_1$ and statistical distances\footnotetext{In Figs \ref{fig:tree},\ref{fig:all} and \ref{fig:perkol}, ${\cal L}_1$ and statistical distances are denoted as Gauss and Mantegna respectively.}. Figs \ref{fig:tree},\ref{fig:all} show the results of graph analysis and Fig \ref{fig:perkol} the time evolution of the percolation threshold for different distance measures. Despite differences in values between results obtained by LMST and CC methods (the graph weight takes its maximal value up to 12 in the case of ${\cal L}_1$ in LMST, whereas in CC  the maximal value of the mean distance is not larger than 1.2) the time evolutions of the measured parameters show that the distances between countries are monotonically decreasing in time whatever the method of analysis. However for the LMST and percolation threshold in ${\cal L}_1$ metrics the evolution is not monotonous. Yet, since the distances between countries are usually decreasing with time this can be interpreted as a proof of a globalization process. A similar conclusion may be obtained by analysing the percolation threshold of G7 countries (Fig \ref{fig:perkol}). However the results depend on the applied distance measures, which are sensitive to different properties of the analysed time series. In the case of $L_1$ and $L_2$ distances the results do not significantly depend on the visualisation method. But in the ${\cal L}_1$ and statistical distances the results are not unique specially in the case of the percolation threshold Fig (\ref{fig:perkol}).


\section{Conclusions}

The correlation between G7 countries has been analysed using different distance functions and various graph methods. Despite the fact that most of the methods allow to observe a globalization like process it is obvious that their sensitivity to observe correlations are different. It seems that the percolation threshold methods is the most sensitive one, since even for $L_1$ and $L_2$ distance functions it reveals different stages of globalization. One can observe that the correlations achieve their highest value in 1990, at well known significant political changes in Europe. Later on, the correlations remain on a   relatively stable level. Analysing the applied distance functions it has been observed that the noise level is the highest in the case of the ${\cal L}_1$ distance since this method is the most sensitive to the length of the data (required for calculating the distribution parameters). However the ${\cal L}_1$ method seems to be the most appropriate, because it compares the distribution functions taking into considerations all properties of the process.

{ \bf Acknowledgement}

This work is partially financially supported by FNRS convention FRFC 2.4590.01. J. M. would like also to thank SUPRATECS for the welcome and hospitality and the organizers of the 3rd Nikkei symposium for financial support received
in order to present the above results. 

\printindex
%\bibliographystyle{apalike}
%\bibliography{nikkei.bib}

\end{document}